%% file: icwsm2019-evolution.tex
\def\year{2019}\relax
\documentclass[letterpaper]{article} 
\usepackage{aaai18}  
\usepackage{times}  
\usepackage{helvet}  
\usepackage{courier}  
\usepackage{url}  
\usepackage{graphicx}  
\usepackage{comment}
\usepackage[tight-spacing=true]{siunitx}

\usepackage{booktabs} 
\usepackage{tikz}
\usepackage[ruled]{algorithm2e}
\usepackage{subcaption}
\usepackage{caption}
\usepackage{flushend}

\usepackage{wrapfig}
\usepackage{amsmath}
\renewcommand{\vec}[1]{\mathbf{#1}}

\usepackage{diagbox}
\usepackage{siunitx}
\usepackage{graphicx}
\usepackage{cleveref} 
\usepackage{url}
\usepackage{lipsum}
\usepackage{multirow}

\usepackage[normalem]{ulem} 
\usepackage{color}

\usepackage[
backend=bibtex,        
sorting=none,
natbib=true,         
style=authoryear-icomp, 
doi=false,            
url=false,            
eprint=false,         
isbn=false,
hyperref=false,        
backref=false,         
maxnames=1,         
maxbibnames=2,    
sorting=nyt
]{biblatex}

\usepackage{xpatch}
\xpatchbibmacro{date+extrayear}{%
  \printtext[parens]%
}{%
  \setunit{\addperiod\space}%
  \printtext%
}{}{}

\DeclareFieldFormat[inproceedings]{title}{#1}
\DeclareFieldFormat[incollection]{title}{#1}
\DeclareFieldFormat[article]{title}{#1}

\DefineBibliographyStrings{english}{%
    page             = {\ifbibliography{}{p\adddot}},
    pages            = {\ifbibliography{}{pp\adddot}},
}

\DefineBibliographyExtras{english}{}

\DeclareNameAlias{sortname}{family-given}
\DeclareNameAlias{default}{last-first}

\begin{document}
\title{Discovering Archetypes to Interpret Evolution of Individual Behavior}
\author{Kanika Narang$^*$,\hspace{1em} Austin Chung$^{*}$,\hspace{1em} Hari Sundaram$^*$,\hspace{1em} Snigdha Chaturvedi$^{\mathsection}$ \\
$^*$ University of Illinois, Urbana-Champaign \hspace{0.3em} $^\mathsection$ University of California, Santa Cruz
 }
 

\maketitle

\begin{abstract}
 In this paper, we aim to discover archetypical patterns of individual evolution in large social networks. In our work, an archetype comprises of \emph{progressive stages} of distinct behavior. We introduce a novel Gaussian Hidden Markov Model (G-HMM) Cluster to identify archetypes of evolutionary patterns. G-HMMs allow for: near limitless behavioral variation; imposing constraints on how individuals can evolve; different evolutionary rates; and are parsimonious.

Our experiments with Academic and StackExchange dataset discover insightful archetypes. We identify four archetypes for researchers : \emph{Steady}, \emph{Diverse}, \emph{Evolving} and \emph{Diffuse}. We observe clear differences in evolution of male and female researchers within the same archetype. Specifically, women and men differ within an archetype (e.g. \emph{Diverse}) in how they start, how they transition and the time spent in mid-career. We also found that the differences in grant income are better explained by the differences in archetype than by differences in gender. For StackOverflow, discovered archetypes could be labeled as: \emph{Experts}, \emph{Seekers}, \emph{Enthusiasts} and \emph{Facilitators}. We have strong quantitative results with competing baselines for activity prediction and perplexity. For future session prediction, the proposed G-HMM cluster model improves by an average of 32\% for different Stack Exchanges and 24\% for Academic dataset. Our model also exhibits lower perplexity than the baselines.
\end{abstract}

\input{Introduction}
\input{RelatedWork}

\input{Model}
\input{Dataset}
\input{Experiment}

\input{QualResults}

\input{limitations}
\input{Conclusion}
\printbibliography


\end{document}

%% file: Introduction.tex
\section{Introduction}
\label{sec:Introduction}

In this paper, we develop models to understand how individuals evolve in large social networks. The problem is important: as individuals interact within the context established by social norms, they gain in experience, and behavioral changes reflect the newfound experience. However, despite a significant focus on community discovery and their evolution in social networks, our understanding of individual evolution is limited (\citet{Yang:2011, McAuley:2013} are some notable exceptions). Understanding evolutionary patterns is useful in a variety of applications: language evolution~\citep{Danescu}; expertise evolution~\citep{McAuley:2013}; journey optimization in digital advertising platforms.
  
\begin{figure}
 \centering
 \small
 \includegraphics[width=0.9\linewidth,height=4.5cm]{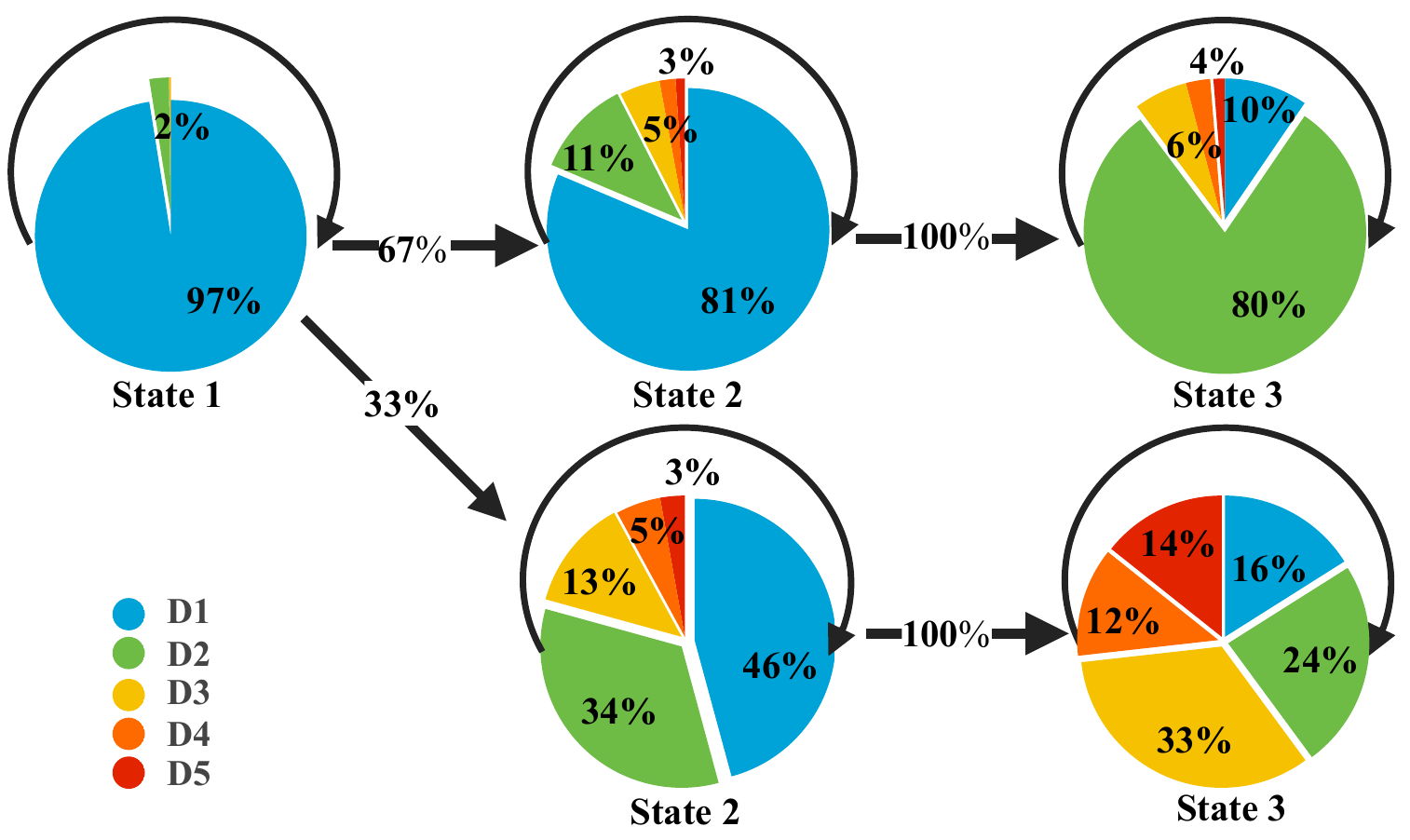}
 \caption{\small \label{fig:example} A stylized academic evolutionary trajectory. Each pie chart is a \emph{behavior stage} in the trajectory. It shows the fraction of papers published in each research area \textit{$D_m$} in that stage. We use a normalized representation focused on change of areas: the label $D_1$ represents the first research area of every academic, $D_2$ the second research area etc. Normalized representations allow us to discover commonalities in behavioral changes of academics across seemingly unconnected domains. In this example, top group of researchers evolve to shift their research focus to a new domain while the bottom group becomes increasingly interdisciplinary.
 }
 \vspace{-0.2in}
\end{figure}

Despite variations in how individuals can evolve, we observe regularities. For instance, for an academic, transition through different stages---PhD Student (focusing on a single research area), being an assistant professor (working on highly related areas) to eventually post-tenure (multiple areas, interests in multidisciplinary collaborations etc.)---mark changes in behavior. These elementary behavioral evolutionary patterns are visible in almost all academic fields, suggesting that surface variations (i.e. area of research for an academic) hide deeper regularities in patterns of behavioral change. We refer to these \textit{latent} regularities in individual behavior as \textit{archetypes} and we plan to explain all individuals' surface variations (the observed research area on which the academic focuses) with a small set of archetypes. ~\Cref{fig:example} shows a stylized example.

Thus a model for learning archetypes needs to: express large observable behavioral variation while exhibiting latent stochastic regularities governing the change of behavior. Furthermore, the model should allow individuals to evolve at different rates. Finally, the results ought to be interpretable in a post-hoc manner.






Our work makes the following contributions:
\begin{description}
\item[A framework for modeling evolutionary trajectories:] We propose a sophisticated framework to identify dominant, interpretable, archetypes amongst individuals for modeling the evolution of their research interests. In contrast, prior work on user modeling has either focused on qualitative analysis (e.g.,~\citep{Ward:2001}), or use engineered features~\citep{Angeletou:2011} or ignored temporal changes ~\citep{Mamykina:2011}. In our work, we assume that an archetype is a \emph{probabilistic model} that encodes individual progression through stages of \emph{distinct} behavior. Specifically, we learn a Gaussian Hidden Markov Model (G-HMM) to capture this progression where latent states capture \emph{behavioral stages} in the evolution. To encode the idea of experience, while we allow individuals to continuously evolve onto the higher stages, we constrain our model to prevent individuals from returning to a stage from which they have evolved. We model \emph{all} individuals with a \emph{small} set of archetypes. We jointly learn the mapping of users into their archetype and the archetype's model parameters through an Expectation-Maximization framework.
\item[Identification of dominant archetypes:] We apply our model to understand the evolution of research interests of Computer Scientists and the users of \emph{Stack Exchange} by studying their activities on the platform. Specifically, for academic dataset, we identify four archetypes: (i) \emph{Steady} researchers who primarily work in their first research area through out their career; (ii) \emph{Evolving} researchers, who continuously shift their dominant area of research; (iii) researchers with \emph{Diverse} research interests; and (iv) researchers who have \emph{Diffused} interests with infrequent contributions in multiple areas. Each archetype is significantly different ($p< .001$) from the others.
\item[Qualitative analysis---variation by gender:] We examined qualitatively, a subset of our data---all full professors (as of Spring 2018) in the top 50 CS departments in United States for gender differences in their academic trajectory. We observe significant differences in the models that explain the evolution of male and female researchers within the same archetype. For example, the models that explain women and men differ (significance level: $p< .01$) in the \emph{diverse} archetype; we observe differences in where they start, rates of transition and research interests during mid-career.
\item[Qualitative analysis---effect on grant income:]  We examine grant income (as of Spring 2018) from the National Science Foundation in US for the same subset of CS academics, to understand the relationship between variations in grant income over the course of academic trajectory and how difference in archetype or gender could serve as explanations. We find significant differences in grant income across genders \textit{within a behavioral stage} of an archetype. For example, for the \textit{steady} archetype, there are differences ($p< .05$) in grant income across genders at the starting stage. Regardless of gender, we also find significant differences \textit{across} behavioral stages within an archetype. For example, we find significant differences ($p< .001$) in grant income between stages 2 and 3 of \textit{all} archetypes.
\end{description}

Additionally, we have strong quantitative results with competing baselines for activity prediction and perplexity on both Academic and Stack Exchange communities. The proposed G-HMM cluster model improves by 24\% for Academic and on an average of 32\% for Stack Exchange communities for future session prediction. Our model exhibits lower perplexity than the baselines. Our model improves by 149\% and 25\% for predicting trajectory of unseen users for Academic and Stack Exchange communities respectively.

\textbf{Significance:} We propose a sophisticated probabilistic framework to identify dominant, interpretable, evolutionary archetypes. We show that the discovered archetypes are significantly different and are straightforward to use to test hypotheses (e.g. evolutionary variation with gender; effects of gender on income). 

We organize the rest of this paper as follows. We discuss related work in Section \ref{sec:related}. In Section \ref{sec:model}, we formally describe our model followed by dataset description in Section \ref{sec:dataset}. We then discuss discovered trajectories in Section \ref{sec:trajectory}. Section \ref{sec:experiments} describes our experiments and its limitations in \Cref{sec:Limitations} and we finally conclude in Section \ref{sec:conclusion}. 

%% file: RelatedWork.tex
\section{Related Work}
\label{sec:related}
\textbf{Activity Modeling}: Our work is most similar to activity sequence modeling that predicts next action or event in a sequence. We are different from those works as our focus is on modeling user \emph{behavior} that is how a user spends her time among possible actions in each session. \citet{Yang:2014} and \citet{Knab2003} proposed generative models that assigns each action to a progression stage and classify event sequences simultaneously. They used their model to predict cancer symptoms or products user would review in the future. However the model did little to provide meaningful and interpretable stages and clusters. The major contribution of our work is in giving an \emph{interpretable} model that helps us to characterize the temporal changes in user \emph{behavior}.

Hidden Markov Model (HMM) have been used to model and cluster time sequences \citep{Smyth:1997,Bicego:2003, Coviello:2014} in the past. However, most of these models learn an HMM for each user sequence and then employ clustering algorithms to cluster the learned HMMs. These approaches are not scalable and the clusters thus identified are not interpretable.

\textbf{User Profiling}: There also has been work in the past on identifying and characterizing user roles in online social networks (OSNs). \citet{Maia:2008} identified five distinct user behaviors of YouTube users based on their individual and social attributes. While \citet{Mamykina:2011} identified user roles based on just answer frequency in StackExchange. Similar user behavioral studies are done by \citep{Adamic:2008} and \citep{Furtado:2013} on Yahoo Answers and Stack Overflow respectively. These studies however ignore \emph{temporal changes} in the behavior and use engineered features for behavior modeling. 

Some behavioral studies do model the evolution too. \citet{Benevenuto:2009} learnt a Markov model to examine transition behavior of users between different activities in Orkut in a static snapshot. \citet{Angeletou:2011} constructed hand crafted rules to identify user roles and study change of user roles' composition in the community over time. Our model, instead, works directly on raw activity data and cluster users with similar pattern of behavioral \emph{evolution}. 

\textbf{Academic Data Mining}: There has been extensive interest in mining Academic Data (bibliographic data, researchers' usage of social media etc.). Studies have been done to understand the evolution of research interests on a community level. \citet{Biryukov:2010} worked on understanding scientific communities in DBLP dataset while \citet{liu2014chi} focused on evolution of research themes in CHI papers over time. \citet{Chakraborty:2018} studied trajectories of successful papers in computer science and physics by analyzing paper citation counts. On an individual level, \citet{Danai:2018} studied career transitions between academia and industry for Computer Science researchers. Studies have also explored usage of Twitter by the Academic community \citep{Maryam:2018, Hadgu:2014, JWS:2017}. In contrast, our focus is on finding commonalities in evolution of \emph{research interests} of scientists across subdomains. 

Recent studies also look at gender differences in funding patterns, productivity and collaboration trends in academia \citep{Way:2016,Way:2017}. Some earlier studies also reported gender differences in academia. \citet{Kahn:1993} identified gendered barriers in obtaining tenure for academics in economics, while \citet{Ward:2001} found gendered differences in pay related to publication record. On the other hand, we explore \emph{gender differences} in a complementary dimension of change in research interests and its effect on grant income.

%% file: Model.tex
\section{Modeling Evolution Trajectories}
\label{sec:model}
In this section, we first describe the model requirements followed by the problem definition. Then, we present our approach that satisfies those requirements.

\subsection{Model Requirements}
An individual's behavior evolves with experience. We operationalize the notion of experience as a \emph{progression through behavioral stages}. A \emph{behavioral stage} is a period of time where individuals exhibits stochastic regularities in their behavior. An individual's \textit{trajectory} is a specific progression through behavioral stages.

We expect individual \emph{evolutionary trajectories} to be unique. However, as the stylized example from the introduction on academic life suggests, despite differences in sub-fields (say HCI vs Data Mining), individuals show latent regularities over time---that is, we observed latent regularities in \emph{how they change}, and \emph{not in what they do.} Latent regularities suggests that we can represent \textit{all} individual trajectories with a finite set of latent, dominant evolutionary trajectories denoted as \emph{archetypes}.
Therefore, models that discover archetypes should satisfy these requirements:
\begin{description}
 \item [Large observed variation:] The model should allow for large (possibly infinite) observed individual behavioral trajectories; each with different behavioral stages.
 \item [Stochastic regularities:] The model should reflect two empirical observations: individuals appear to progress through a series of \textit{distinct} behavioral stages reflecting gain in experience; individuals exhibit regularities in how they evolve through these stages.
 \item [Different rates of evolution:] Individuals evolve at different rates and can skip stages.
 \item [Parsimonious:] We should be able to approximate individual trajectories by small number of \emph{dominant} trajectories.
\end{description}


We want the model to be post-hoc interpretable~\citep{Lipton2016}. That is, we would like model outcomes, including the definition of stages and how individuals progress amongst them to allow for meaningful interpretation (e.g. `the person started to work on multidisciplinary research after tenure').

\subsection{Problem Definition}
Now, we discuss the problem formally. We represent an individual $i$, as a time ordered sequence, $\mathbf{X_i}$, of their activities, over a time granularity appropriate to that domain. For example, for the Stack Exchange data, the granularity is that of a single visit to the network, while in case of the Academic corpus, the granularity is that of a year since most conferences occur annually.

Without loss of generality, we refer to the fundamental temporal unit of analysis as a \emph{session}. Thus, $\mathbf{X_i}$ is a sequence of sessions, $\vec{X}_{ij}$, where $j \in \{1, 2, \ldots t_i\}$ and $t_i$ is the number of sessions for an individual $i$. In general, lengths of sequences will vary across individuals depending on their activity level. A session, $\vec{X}_{ij}$, is a vector $\langle o_1, o_2, \ldots, o_M \rangle$, where $M$ is the number of different actions possible in that network. Each element $o_m$ of the vector $\vec{X}_{ij}$, denotes the fraction of time the individual performs the $m$-th action during a single session. For example, for the Stack Exchange data, if the set of possible actions include `posting a question', `answering a question', and `commenting on answers or on questions', then a session is a distribution over these three actions during a single visit to the social network.

The problem addressed in this paper is to associate an \emph{archetype} with each individual.
We assume that there exist $C$ different archetypes, and given a sequence of sessions for an individual $\vec{X}_i = \{\vec{X}_{i 1}, \vec{X}_{i 2} \ldots \vec{X}_{i t_i} \}$, the goal is to assign the sequence to one of the $C$ archetypes---each associated with a set of $K$ latent behavioral stages. During this assignment, we also identify how the individual evolves through its archetype's distinct stages by outputting the sequence $Y_i = \{Y_{i 1}, Y_{i 2} \ldots Y_{i t_i} \}$, where $Y_{i j}$ represents the behavioral stage $k \in [1,K] $ assigned to $j$-th session in individual $i$'s evolutionary trajectory. We constrain the number of stages $ K \ll t_i$, and allow skipping of stages, while disallowing return to earlier stages.

\subsection{A Framework for Identifying Archetypes}
\label{subsec:GHMMCluster}
We use a Gaussian-Hidden Markov Model (G-HMM) based approach to model individual behavior.
To capture broad variations amongst individuals, we learn a set of  $C$ G-HMMs where each G-HMM represents an archetype. We jointly learn the partitioning of the individuals into different archetypes and the model parameters for each archetype.





Each Gaussian HMM, associated with an archetype $c$, has $K$ discrete latent states and $\pi^{c}$ is a $K$ dimensional prior vector for its latent states. The model makes a first order Markovian assumption between state transitions using the transition probability matrix $\mathbf{{\tau}^{c}}$; where $\tau_{kl}^{c}$ represents the probability of transitioning from state $k$ to $l$ in the $c$-th archetype. Lastly, the model assumes that given a latent state, $k$, from an archetype $c$, the $M$ dimensional session vector, $X_{ij}$, is Normally distributed with mean $\mu_{k}^{c}$ and covariance $\mathbf{{\Sigma}_k^{c}}$.

In the above generative process, the G-HMM associated with different archetypes do not share latent states. In other words, each G-HMM has its own set of discrete latent states.\footnote{ Experiments with tied-states of archetypes led to worse results.}However, we fix the number of states ($K$) to be the same for each archetype.

\textbf{Encoding Experience \& Variable Evolutionary Rates: }
To encode the idea of experience, as well as to allow variable evolutionary rates, similar to~\cite{Yang:2014}, we allow only forward state transitions (including self loop) within a G-HMM that represents an archetype. This choice appears sensible to us since semantically, each latent state of the G-HMM represents a \emph{behavioral stage} of evolution, and its corresponding mean vector encapsulates \emph{behavior} distribution in that stage. Then, forward transition denotes \emph{progression} through \emph{behavioral stages}. We operationalize this idea by using an upper triangular state transition matrix.

\textbf{Training:} We train our Gaussian HMM archetype model using a (hard) Expectation Maximization~\citep{Dempster:1977} based iterative procedure described in Algorithm~\ref{euclid}. During training, the goal is to learn the G-HMM parameters, $ \mathbf{\lambda^c}$, for each archetype $c$, where $\mathbf{\lambda^c} = \langle\mu^c, \mathbf{\Sigma^c}, \pi^c, \tau^c \rangle$ and archetype assignments for each user, $c_i$. We begin with initializing the Gaussian HMMs with initial parameters, $ \mathbf{\lambda_0^1}, \mathbf{\lambda_0^2}, \ldots, \mathbf{\lambda_0^C}$. Thereafter, in the iterative training process, in the Expectation step, we use current estimates of $\mathbf{\lambda^c}$ 's to assign an archetype to each user sequence in the data. In the Maximization step, we use current archetype assignments to learn the corresponding G-HMM's parameters $\mathbf{\lambda^c}$. We use a modified version of the Baum-Welch algorithm~\citep{Rabiner:1990} allowing for forward-only transitions. Thus, this method jointly partitions the input sequences into different archetypes as well as learns the parameters of the associated G-HMMs.

\begin{algorithm}[tbh]
  \small
 \caption{Gaussian HMM archetype}\label{euclid}
 \SetAlgoLined
 \textbf{Input:} $\vec{X_i}$ and $\mathbf{\lambda^c_0}$ $\forall i \in \{1, 2, \ldots N\}$ $\forall c \in \{1, 2, \ldots C\}$\;
 \textbf{Output:} $\vec{Y_i}$ and $\mathbf{\lambda^c}$ $\forall i \in \{1, 2, \ldots N\}$ $\forall c \in \{1, 2, \ldots C\}$\;
 Initialize the $c^{th}$ archetype with initial parameters, $\mathbf{\lambda^c_0}$ $\forall c$\;
 \While{ not converged} {
  \textbf{M-Step:} Re-assign archetypes to sequences $\mathbf{X_i}$ as: \\
  $c_i$ =  $argmax_{c} P(\mathbf{X_i} | \mathbf{\lambda^c})$ $\forall i \in \{1, 2, \ldots N\}$\;
  \textbf{E-Step:} Re-estimate the G-HMM parameters, $\mathbf{\lambda^c} \forall c \in \{1, 2, \ldots C\}$, using modified Baum-Welch algorithm.\;
 }

 \textbf {Convergence Criteria}\;
 \begin{itemize}
   \itemsep0em
  \item Log Likelihood difference falls below threshold; or\
  \item Number of iterations is greater than threshold; or\
  \item Number of sequences re-assigned in an iteration is less than 1\% of the data\
 \end{itemize}
\end{algorithm}

\textbf{Implementation Details: }
Our iterative training procedure requires initialization for G-HMM parameters $\mathbf{\lambda^c_0}$. We  perform k-means clustering on all sessions of all user sequences in our corpus, treating the sessions as independent of each other (thus losing the sequential information). The cluster centers, thus obtained are used as the initial means, $\mu^c_0$, for the latent states. We fix each $\Sigma^c_k$ as an identical diagonal covariance matrix $\sigma I$ with $\sigma = 0.01$ based on preliminary experiments. We initialize transition matrices, $\tau^c_0$, and states' prior probabilities, $\pi^c_0$, for each archetype randomly.

Our implementation is based on Kevin Murphy's HMM Matlab toolbox~\footnote{\url{bit.ly/hmmtoolbox}}. Also, we implement a parallelized version of our EM algorithm to reduce computation time. We test our model on Intel Xeon Processor with 128 Gb RAM and clock speed of 2.5 GHz. Our model takes around 10 minutes on our biggest dataset Stack Overflow (570K users) using parallelization with 30 threads.

%% file: Dataset.tex
\section{Dataset}
\label{sec:dataset}
For our analysis, we use the Microsoft Academic Dataset and Stack Exchange Dataset. Table \ref{tab:stats} shows the data statistics, and we provide a brief description of the two corpora in the rest of this section.

\begin{table}[tbh]
 \centering

 \small
 \begin{tabular}{p{17mm} r r r r}
  \toprule
  Dataset                 & N       & $\bar{t}$ & $t_{\max}$ & M \\  \midrule
  Academic       & 4578      & 24.15   & 47     & 6                      \\
  StackOverflow & 561937    & 47.13   & 750    & 5                      \\
  English & 3828    & 44.01   & 729    & 5                      \\
  Money & 873   & 44.41   & 706    & 5                      \\
  Movies & 678    & 48.40   & 598    & 5                      \\
  CrossValidated & 3728    & 38.94   & 738    & 5                      \\
  Travel & 1000    & 56.14   & 736    & 5                      \\
  Law & 195    & 47.79   & 584    & 5                      \\
   \bottomrule
 \end{tabular}
 \caption{ \footnotesize \label{tab:stats}Dataset statistics for the Academic and Stack Exchange datasets. $N$: number of users; $M$: possible actions in each session; $t_{\max}$: maximum session length; $\bar{t}$: mean session length. For authors, $\bar{t}$ is their average career length (in years). }
 \vspace{-0.2in}
\end{table}

\subsection{Microsoft Academic Dataset}
We use the Microsoft Academic dataset to study evolutionary patterns of researchers with a focus on Computer Scientists. To this end, we extract publication history of authors in Computer Science (CS) using the Microsoft Academic Knowledge Service API \footnote{\url{http://bit.ly/microsoft-data}}. Microsoft Academic Service additionally annotates each publication with the year of publication, publication venue and the CS subfield (out of $35$ identified fields) to which it belongs. 

For this study, we decide to focus only on \emph{influential} scientists with sufficient publication history. We identify \emph{influential} authors based on \emph{prominence} of the conference venues in which they publish. To quantify \emph{prominence} of a conference, we construct a conference-conference citation graph where each conference in our dataset forms a node and the weighted edges represent inter-conference citation frequency. Specifically, the weight of a directed edge from conference $C_1$ to conference $C_2$ is proportional to the fraction of papers published in $C_2$ cited by papers published in $C_1$. We then use the Pagerank algorithm \citep{ilprints422} on this directed graph and define conference \emph{prominence} as the Pagerank of the corresponding conference-node. Thereafter, we define an author's \emph{influence} as the weighted sum of prominences of the conferences (s)he has published in. Here, conference-prominences are weighted by the fraction of the author's papers published in that venue. 

We rank authors in decreasing order of their \emph{influence} and extract top $750$ most-influential authors from each of the $35$ CS areas in the dataset. Note that authors can be \emph{influential} in more than one subfield. We then extract unique authors from this set who have at least 15 years of publication history. We only consider publication history from 1970 to 2016 to avoid missing data. The resulting dataset consists of records of $4578$ authors.\footnote{This data will be made available upon publication}

We now describe how we represent an author's academic life-cycle as a sequence, $\mathbf{X_i}$, comprising of session-vectors, $X_{ij}$. We chose each session to be an year long as most CS conferences occur annually. For this dataset, a session-vector represents the fraction of papers an author publishes in various \emph{area-of-interests} (\texttt{AoI}s) in that year.

For defining an \texttt{AoI} of an author, we consider all papers published by the author in her academic life. We identify her primary \texttt{AoI}, $D_1$, as the \emph{first} subfield (out of 35 subfields) in which she publishes \emph{cumulatively} at least $3$ papers in the first 3 years. Usually, an author's $D_1$ is about their PhD dissertation work and we expect students to \emph{settle} down after a few years. Thus, after identification of $D_1$, hopefully with a steady paper count, we define her secondary \texttt{AoI}, $D_2$, as the subfield in which she publishes at least $3$ papers in \emph{one} year. Similarly, we also define tertiary ($D_3$), quaternary ($D_4$), and quinary ($D_5$) \texttt{AoI}. 
We do not define \texttt{AoI}s beyond $D_5$ because 80\% of authors do not explore more than $5$ subfields in our dataset. Also, in a given year, if an author publishes fewer than $3$ papers in an unexplored subfield, these papers count towards a sixth dimension \texttt{AoI} called \emph{Explore}. This denotes that the author has started exploring new subfields but they are not yet significant enough to be one of the $D_m$'s ($m \in {[1,5]})$, and indicate a possible shift in research interests. To summarize, each session is a $6$ dimensional vector ($M=6$), and its elements are the fraction of the author's publications in one of the $5$ $D_m$'s or the $6^{th}$ \emph{Explore} dimension. This normalized representation for sessions allows our model to discover behavioral patterns of author's changing research interests in a domain independent manner.


\subsection{Stack Exchange Dataset}
Our second dataset consists of activity logs of users of Stack Exchange~\footnote{\url{https://data.stackexchange.com/}}(as of Feb 2017), a popular online question-answering platform.
In this paper, we work on 7 diverse communities of the platform: 
Stack Overflow, English, Money, Movies, CrossValidated, Travel and Law. These communities have varied sizes and cater to different audiences. For each user, the data contains details about their activities on the community. Stack Exchange allows $5$ different activities ($M=5$): post a \textbf{Q}uestion; \textbf{A}nswer
a question; 
\textbf{C}omment on a question or an answer; \textbf{E}dit operations like assign tags, edit body or title of a post; and \textbf{M}oderator operations like voting. Like before, we represent a user by a sequence, $\mathbf{X_i}$, of session vectors, $X_{ij}$. 
We split the activity-sequence of a user into sessions using a time threshold similar to session definitions in web search~\citep{Narang:2017}. Specifically, we create a new session if the difference between two consecutive activities is more than 6 hours. A gap longer than this 
marks 
a new visit to the community. Hence, a session is a subsequence of the user's activity-sequence and is formally represented as a distribution over the $M$
possible activities; where its $m^{th}$ element represents the fraction of total activity spent in the $m^{th}$ activity in that session.

Lastly, to focus on users who have spent enough time in the network to exhibit behavioral changes, we filter users with less than $10$ sessions, and also remove outliers with more than $750$ sessions.

%% file: Experiment.tex
\section{Quantitative Experiments}
\label{sec:experiments}
In this section, we evaluate our model on two different tasks: Future Prediction and Perplexity. We describe the baselines in Section~\ref{sec:baseline} and report results in Section~\ref{sec:tasks}.

\subsection{Baselines}
\label{sec:baseline}
\textbf{Distance GHMM} : Our first baseline uses the GHMM clustering model as defined in \cite{HMM2014}. In this baseline, we learn a GHMM for each user and then cluster the models using distance metric $\delta$, the symmetric KL divergence ($d_{kl}$) between two G-HMMs~\citep{rainier}.
{\small
\begin{align}
	\label{eq:KL}
	d_{kl} (\lambda^p, \lambda^q) = \frac{1}{N_p} \sum_{i \in N_p} log \frac{P(X_i | \lambda^p)} {P(X_i | \lambda^q)},
\end{align}
}%

We use k-medoids clustering; since this method doesn't give a representative model for each cluster, we learn a GHMM per cluster.
For fair comparison, we set $k$, the number of clusters to be the same as our model.

\textbf{Vector AutoRegressive Model (VAR)}: VAR models are used to model multivariate time series data \citep{Ltkepohl:2007}. It assumes that each variable in the vector is a linear function of it's own past values as well as other variables. For each user sequence $\mathbf{X_i}$, $j$th session is modeled as,
{\small 
\begin{align}
  \vec{X}_{ij} = A_1 \vec{X}_{ij-1} + \ldots + A_p \vec{X}_{ij-p} + u_j
\end{align}
}%
where $A_i$ is \emph{M} X \emph{M} matrix, $u_j \sim \mathcal{N}(0,\,\Sigma_{u}) $ and we set $p=1$ as in first-order Markov models.

\textbf{Gaussian clusters (GCluster)}: In this baseline, we assume that individuals \emph{do not evolve} in their lifespan. This is a simplified version of our model. It assumes that there are different archetypes but that each archetype has only one state. Hence, all sessions of a sequence are generated from a single  multivariate Gaussian.

We can not compare with other 
sequence prediction baselines~\citep{Yang:2014, Knab2003} as they assume a discrete set of activities while in our case, each session is a probability distribution over possible activities.

\subsection{Tasks}
\label{sec:tasks}
\textbf{Future Prediction}:
In this task, we predict future behavior of an individual given her history. We assign the first $90\%$ sessions of each sequence for training, and predict the behavior in future sessions (the remaining $10\%$ of the sequence). We first use all the training sessions to learn parameters of our model. Then, for each sequence, we run Viterbi algorithm to decode state assignment of its test sessions, \emph{$t^\prime_i$}. The test sessions of the $i$-th user will have same archetype assignment $c_i$ determined in the training session for that user.

We compute Jensen-Shannon($d_{js}$) divergence between the mean $\mu^{c_{ij}}$ of the assigned state $Y_{ij}$ and the observed vector $X_{ij}$. $d_{js}$ is a symmetric K-L divergence between two vectors. We report the average $\bar{\Delta}$ over all test sessions:
{\small 
\begin{align}
	\label{eq:pred}
  \bar{\Delta} = \frac{1}{\textit{$|T|$}} \sum_{i \in N, j \in t^\prime_i} d_{js}(\mu^{c_{ij}}, X_{ij}), \\
  d_{js}(\mu^{c_{ij}}, X_{ij}) = \frac{1}{2}d_{kl}(\mu^{c_{ij}}, p) + \frac{1}{2}d_{kl}(X_{ij}, p),
\end{align}
}%
where, $p = \frac{1}{2}(\mu^{c_{ij}} + X_{ij})$ and $d_{kl}$ measures KL divergence distance. For VAR, we use the model learnt on training sessions of  user $i$ to make prediction for her future sessions.

Table \ref{tab:futurepred} shows our results on this task for different datasets. Our model outperforms the baselines for all Stack Exchange datasets with an average improvement of about 32\% and 24\% on the Academic dataset. Hence, learning archetypes can help us to accurately predict an individual's future behavior in the social network.

\begin{table}[tbh]
	\centering
\sisetup{
  round-mode = places,
round-precision = 2
}
		\small 
		\begin{tabular}{lSSSSS} \toprule%
			Dataset                & {Our Model}& {VAR}& {\shortstack{Distance \\ HMM}} & {\shortstack{Gaussian \\ Cluster}} \\ \midrule 
			Academic               & 0.2189  &  0.31    &   0.4151  &  0.2941    \\ 
			{StackOverflow}       & 0.2289  &   0.3577  & NA  &  0.3715     \\
			{English}              & 0.187 & 0.29 &  0.26      & 0.31 \\
			Money                  & 0.186 & 0.52 &  0.32      & 0.32  \\
			Movies                 & 0.23  & 0.35 &  0.3502    & 0.37 \\
			CrossValidated         & 0.21  & 0.38 &  0.3260    & 0.35  \\
			Travel                 & 0.19  & 0.30 &  0.2547    & 0.29  \\
			Law                    & 0.19  & 0.26 &  0.33      & 0.27  \\ \bottomrule
		\end{tabular}
	\caption{ \small \label{tab:futurepred} Average Jensen-Shannon divergence of future sessions using 90-10\% split of each user sequence. Lower values are better. Distance HMM did not converge on StackOverflow dataset. 
   }
\end{table}

\textbf{Perplexity}
Perplexity measures how surprised the model is on observing an unseen user sequence. A lower value of perplexity indicates low surprise and hence a better model.
{\small 
\begin{equation}
	P_x = - \frac{1}{ \textit{$|T|$} } \sum_{i \in T }  \max_{c \in C}  (\log P(\mathbf{X_i^T} | \lambda^{c}))
\end{equation}
}%
where, $\mathbf{X_i^T}$ represents a test sequence in Test Set \emph{T}, and $\lambda_c$ represents the parameters of the GHMM corresponding to the $c$-th archetype. We assign $\mathbf{X_i^T}$ to the archetype \emph{c} with maximum likelihood. Perplexity is then computed as the average likelihood of all test sequences.

Table \ref{tab:perplexity} reports average perplexity after five fold cross validation. Note that for this experiment, model predicts entire trajectory of a new user. We could not use the regression baseline (VAR) as it is not a generative model. Our model beats best performing baseline by 149\% on Academic and by around 25\% on average for StackExchange datasets. Hence, our model also effectively predicts behavior of future individuals joining the social network.

\begin{table}[tbh]
	\centering
	\sisetup{
		round-mode = places,
		round-precision = 2,
	}
	\small 
		\begin{tabular}{lSSSS}
			\toprule %
			Dataset         & {Our Model}     & {\shortstack{Distance \\ HMM}} & {\shortstack{Gaussian \\ Cluster}}  \\ \midrule
			Academic        & -18.37  & 37.73  & 100.79       \\ 
			StackOverflow   & 487.68  & NA           & 678.62   \\
			English          & 306.38  & 559.6459   & 471.137  \\
			Money           & 415.853 & 557.686  & 570.509   \\
			Movies          & 596.10  & 724.15   & 743.73    \\
			CrossValidated  & 398.442 & 514.7365  & 554.313   \\
			Travel          & 494.061 & 645.6434  & 666.966  \\
			Law             & 368.894 & 508.077  & 482.267 \\ \bottomrule
		\end{tabular}
	\caption{\small \label{tab:perplexity} Average Perplexity on unseen user sequences after 5 fold cross validation. Lower values are better. DistanceHMM did not converge on StackOverflow dataset. 
  }
\end{table}

\textbf{Discussion:} For future prediction, our model performs better than VAR model. It shows that modeling cluster of sequences gives a better estimate than modeling each user sequence separately. Also, if we assume no behavior evolution and just cluster users according to their behavior i.e. \emph{GCluster} model, we obtain worse results.
 Our model also outperforms similarity distance based clustering method: DistanceHMM \citep{HMM2014}, which is also the strongest baseline. It first estimates G-HMM model for each user sequence and then cluster these models. Estimating model for each sequence can be noisy, specially if the user sequence has short length. Instead, when we jointly learn G-HMM model parameters and cluster sequences, we learn a better approximation.

\textbf{Full vs. Left-Right Transition Matrix}: We also test our model with unconstrained full transition matrix where users can jump from a state to any other state in the HMM. We compare our results with this model for future prediction task. On academic dataset, we obtain similar values while for StackExchange communities, the full transition matrix gives better results. This can be because a full matrix has more degrees of freedom but then, it is also more expensive to learn. Also, the states learnt are not interpretable. As \citep{Yang:2014} and \citep{Knab2003} noted, forward state transitions accurately models the natural progression of evolution, we thus, chose to work with a forward transition matrix.

%% file: QualResults.tex
\section{Qualitative Analysis}
\label{sec:trajectory}
In this section, we perform qualitative analysis of archetypes identified by our model. Due to space constraints, we analyze the archetypes discovered by our model only for the Academic Dataset and Stack Overflow (programming based Q\&A); largest and most popular community in StackExchange. We first describe the discovered archetypes of all researchers in~\Cref{sec:acad}. Then, we examine gender variation in academic trajectory in~\Cref{sec:discussion} and effect of archetype and gender on grant income in~\Cref{sec:grant}. Finally, we details archetypes for Stack Overflow in ~\Cref{sec:stack}.


\subsection{Academic Archetypes}
\label{sec:acad}
Our analysis reveals four archetypes: \emph{Steady}, \emph{Diverse}, \emph{Evolving} and \emph{Diffuse}. We chose the number of clusters $C=4$ using the elbow method \cite{elbow:2001}: data log likelihoods increased rapidly till four clusters with much slower increase beyond that. Further, we chose number of states per cluster, $K=5$: beyond five states, KL divergence\cite{kl:1951} between mean vectors of new states with previous states started reducing rapidly indicating redundant states.

We also conducted t-test to validate differences among the identified archetypes. Specifically, paired-sample t-test \cite{goulden:1949} is conducted between likelihood values of data points assigned to an archetype with their likelihood values obtained from rest of the archetypes. For instance, for each archetype pair $(p, q)$, we conduct paired t-test between $\log P(X_i| \lambda^p)$ and $\log P(X_i| \lambda^q)$ $\forall i \ni c_i=p$. Note that test results for archetype pair $(p, q)$ are not symmetric.
We observed that all archetype pairs are significantly different ($p < .001$). Now, we first discuss what is common to these discovered archetypes before examining each one in detail.

\begin{figure*}[tbh]
	\centering
	\begin{subfigure}{\textwidth}
		\centering
		\includegraphics[width=\linewidth,height=6cm]{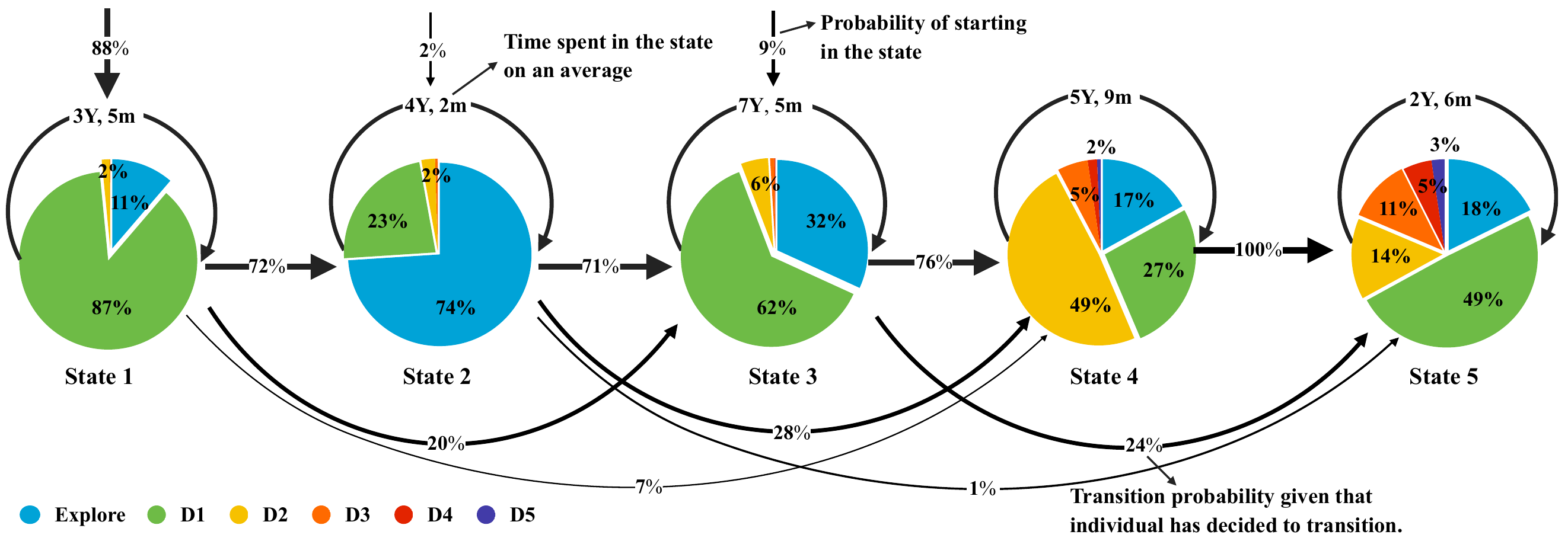}
	\end{subfigure}
	\caption{\small \label{fig:acadclusters}  Trajectory (state sequence) for \emph{Steady} archetype in the Academic Dataset. Each pie is a \emph{latent state} or \emph{behavior stage} in the trajectory. It denotes the proportion of papers published in each \emph{Area of Interest}'s in the latent state. Each state is also labeled with the average amount of time spent in the state. For example, in this cluster, 87\% of publications in the first 3.5 years are in author's primary \texttt{AoI} $D_1$ while rest 11\% are in exploring other areas. The arrows on the top of each pie show prior probability for starting in that state. As we learn a left-to-right G-HMM, author can transition to its immediate next state or any later latent states. Each transition is labeled with the corresponding conditional transition probability i.e. transition probability given that the user has decided to transition. The arrows thickness is proportional to it's weight. Authors in this cluster exhibit \emph{steady} research interest in their primary \texttt{AoI} $D_1$. Some authors start dominantly contributing in their secondary \texttt{AoI}, $D_2$ in State 4. Though, they return to spending around half of their effort in $D_1$ in State 5.		}
	\label{fig:academic}
\end{figure*}

\begin{table}
	\footnotesize
	\begin{tabular}{@{}l p{15mm} p{15mm} p{15mm} p{15mm}@{}}
		\toprule
		  & Steady & Diverse & Evolving & Diffuse \\
		\midrule
		State 1& \{3Y, 5m\} \newline $\mathbf{D_1}$ (87\%) \newline Ex (11\%) &
		\{3Y, 3m\} \newline $\mathbf{D_1}$ (88\%)\newline Ex (11\%)&
		\{2Y, 9m\} \newline $\mathbf{D_1}$ (72\%)\newline Ex (24\%) &
		\{2Y, 7m\} \newline $\mathbf{D_1}$(76\%) \newline Ex (22\%) \\
		State 2 &  \{4Y, 2m\} \newline \textbf{Ex} (74\%)\newline ${D_1}$ (23\%) &
		\{2Y, 6m \} \newline \textbf{Ex} (80\%)\newline ${D_1}$ (16\%) &
		\{2Y, 9m \} \newline \textbf{Ex} (83\%)\newline ${D_1}$ (12\%) &
		\{3Y, 7m \} \newline \textbf{Ex} (91\%) \\
		State 3 & \{7Y, 5m\} \newline $\mathbf{D_1}$ (62\%) \newline \textbf{Ex} (32\%)  &
		\{5Y, 6m\} \newline $\mathbf{D_1}$ (73\%) \newline Ex (17\%) &
		\{6Y, 2m\} \newline $\mathbf{D_1}$ (33\%) \newline \textbf{Ex} (28\%) \newline $\mathbf{D_2}$ (24\%)&
		\{8Y, 5m\} \newline $\mathbf{D_1}$ (50\%) \newline \textbf{Ex} (39\%) \\
		State 4 & \{5Y, 9m\} \newline $\mathbf{D_2}$ (49\%) \newline $\mathbf{D_1}$ (27\%) \newline Ex (17\%) &
		\{5Y, 6m\} \newline $\mathbf{Ex}$ (46\%) \newline ${D_2}$ (20\%) \newline ${D_1}$ (17\%)&
		\{5Y\} \newline $\mathbf{D_2}$ (66\%) \newline $Ex$ (18\%)  &
		\{3Y, 9m\} \newline $\mathbf{D_2}$ (43\%) \newline $\mathbf{Ex}$ (26\%) \\
		State 5 & \{2Y, 6m\} \newline $\mathbf{D_1}$ (49\%) \newline {Ex} (18\%) \newline ${D_2}$ (14\%) &
		\{6Y, 3m\} \newline $\mathbf{D_4}$ (29\%) \newline \textbf{Ex} (20\%) \newline ${D_3}$ (14\%) \newline ${D_1}$ (14\%) &
		\{6Y, 5m\} \newline $\mathbf{D_3}$ (43\%) \newline {Ex} (19\%) \newline ${D_2}$ (14\%) &
		\{4Y, 1m\} \newline \textbf{Ex} (74\%) \\
		\bottomrule
	\end{tabular}
	\caption{ \small \label{tab:mean} Learned mean vector for each state for four archetypes in the Academic Dataset. We list the \emph{Area-of-Interests} (\texttt{AoI}) in sorted order and annotate them with their \% contribution in the state. We list main \texttt{AoI} ($>$ 11\%) for each state. Each state is also labeled with it's average duration in \{Years (Y), months (m)\}. The labels given to these clusters reflect our own interpretation of the user behavior and make disambiguating the behavior easier in the text.}
\end{table}

\begin{table}[tbh]
	\centering
	\small
	\begin{tabular}{p{28mm} r r r r} 
		\toprule
		        & Steady  & Diverse & Evolving & Diffuse \\ \midrule
		Male    Professors  & 247     & 206     & 241    & 263   \\
		(Top-50 US schools) &      &      &     &    \\
		Female Professors (Top-50 US schools) & 30     & 32    & 26    & 39  \\ \hline
		Total Authors   & 1329    & 1080     & 1107    & 1062   \\ \bottomrule %
	\end{tabular}
	\caption{ \small \label{tab:acadclusterdata} Statistics for discovered archetypes. Identified Male and Female professors in Top-50 US schools. Total authors includes all researchers and professors in the dataset.}
\end{table}

\texttt{Commonalities in Archetypes:} Table \ref{tab:mean} summarizes the trajectories (state sequences) learned for the four different archetypes in this dataset. Each archetype is labeled according to our own interpretation of the user behavior, looking at the learned mean vector of GHMM states. We observe that all archetypes exhibit similarities, especially in the first two stages. Across all archetypes, the first \emph{stage} typically spans around $3$ years, and more than 72\% of the published research is in the author's \emph{primary} \texttt{AoI}: $D_1$. As noted before, this is most likely their PhD dissertation area and hence, the research is more focussed. After gaining some research experience, most authors move to the second \emph{stage} where they start exploring other research areas denoted by marked increase in their \emph{Explore} \texttt{AoI}(more than 74\%). However, in state 3 and beyond, authors from different archetypes follow different trajectories where they differ in how they change their dominant \texttt{AoI} over time while \emph{exploring} other domains. Below, we describe each of the four trajectories in more details.

\texttt{Steady}: The first major archetype is of \emph{steady} researchers, who mainly work in \emph{one} \texttt{AoI} (i.e. their $D_1$) throughout their career. Fig \ref{fig:academic} shows the states of this archetype. We can see that most people start in their primary \texttt{AoI}, $D_1$ (state 1), which possibly reflects their PhD education. After graduation, they spend some time \emph{exploring} other areas while continuing to publish in $D_1$ (state 2), but move back to publishing in $D_1$ for a significant portion of their careers, about 7.5 years (state 3). This is often again followed by a phase where they start working in another area, $D_2$, while continuing to publish in $D_1$ (state 4), they eventually revert to publishing in $D_1$ (state 5) towards the latter part of their careers. In the last state, they also publish widely in other areas (indicated by almost half of the pie divided between other $D_m$'s), but their main interest remains $D_1$.
Michael Jordan, professor at University of California, Berkeley exhibits this research trajectory. He is a Machine Learning expert; his primary \texttt{AoI} $D_1$, and has secondary interests in Data Mining, Optimization and Bioinformatics.
Theory professor at University of Illinois, Urbana-Champaign, Jeff Erickson is also assigned to this cluster; he also publishes in his primary \texttt{AoI} $D_1$ (Theory) with auxiliary interests in the field of mathematical optimization.

\texttt{Diverse}: The second archetype consists of researchers with \emph{diverse} research interests as they make significant contributions in multiple $D_m$'s. Similar to \emph{steady} researchers, they research in their primary \texttt{AoI} $D_1$ while \emph{exploring} other domains in the initial $3$ states as shown in Table \ref{tab:mean}. They, then, publish in $D_2$ and $D_1$ while spending half time \emph{exploring} other possible interests (state 4). They evolve to have strong research presence in all $5$ AoIs (state 5). This behavior suggests that authors of this archetype tend to work in interdisciplinary areas; or projects with broader scope which gain acceptance by different research communities.
One notable example is Prof. Jiawei Han at University of Illinois, Urbana-Champaign, who started his academic career studying Databases and Data Mining, also making significant contributions in Machine Learning and Bioinformatics lately.
Another professor who started in Databases, Jaideep Srivastava of the University of Maryland, evolved on to research distributed implementation of databases, and also data mining and AI related research simultaneously.

\texttt{Evolving}: These researchers have one dominant area of interest (\texttt{AoI}) in each state which \emph{changes} with time. Their dominant  area of interest (\texttt{AoI}) \emph{evolves} from $D_1$ (72\%) in state 1 to $D_2$ (66\%) in state 4 to $D_3$(43\%) in state 5. Even though their \texttt{AoI} shifts across stages, in any given stage, they remain focused on one area and do not publish much in other areas.
James Foley, professor in Georgia Tech, started in Computer Graphics and later switched to research on user-computer interfaces and recently, User Modeling.
Natural Language Processing (NLP) expert Daniel Jurafsky at Stanford University, also steadily moved from pure NLP based research problems to Speech processing, and later to Machine Learning (ML). Also note, for Jurafsky, this evolution can be attributed to the broader shift of using sophisticated ML models to solve NLP problems.

\texttt{Diffuse}: Authors of this archetype stay focussed in one dominant area in each stage; while in the last stage their research interests are \emph{diffused}. Authors publish considerably in one dominant area in first 3 stages; $D_1$ (state 1, 3) to $D_2$ (state 4). In the last state, which lasts around 4 years, the authors are infrequently publishing (less than 3 papers a year) in new subfields accounting for 74\% of their publications. Hence, these authors have \emph{diffused} research interests after they gain experience.
Gerhard Weikum, professor at MPI Germany started in Databases area made a brief transition to Information Retrieval work and later started publishing in Machine Learning and Data Mining fields too. These area evolutions are more natural transitions as they are highly interrelated which explains contributions in all fields.
Anind Dey is a professor at Carnegie Mellon University who initially worked on sensor technology and then switched to Web mining and Human Computing related research problems.

\input{Discussion}

\subsection{StackOverflow Archetypes}
\label{sec:stack}
We now describe archetypes learned for the Stack Overflow data. Table \ref{tab:stackexchangemean} depicts the latent states for all archetypes. We label the 4 archetypes discovered as \emph{Expert}, \emph{Seekers}, \emph{Enthusiasts} and \emph{Facilitators}. Posting Comments is the most frequent activity in all archetypes as it is a very low cost activity. Moderator actions and Edits are least favored activities by Stack Overflow users. Most of the users spend initial sessions for \emph{posting questions} (state 1) and significant proportion of their later sessions in \emph{posting answers or comments} (state 6).

\texttt{Experts} users join the community to answer queries or post clarifications or edit answers (state 1-6). They spend at least 68\% of their sessions in \emph{posting answers} (state 1 \& 3). They rarely ask questions of their own. In 
communities like Stack Overflow, it is vital to have a dedicated group of experts answering queries for it to be sustainable.

\texttt{Information Seekers} 
join the community for getting answer to their queries accounting for 69\% of activities per session (state 1). They briefly start contributing by posting answers to the community (state 3) but they end up again in \emph{commenting} (state 6) or \emph{asking questions} (state 5).

\texttt{Enthusiasts} start by asking questions and posting comments (state 1). They, then, start answering questions and commenting on other answers (state 2). They briefly stay (4 sessions) in edit state (state 3) but end up migrating to either commenting again (state 4) or asking questions and commenting (state 5). We denote them as \emph{Enthusiasts} as they use the platform to post questions while simultaneously answering queries from their acquired knowledge.

\texttt{Facilitators} 
join for information seeking (state 1) but start posting answers, clarifying and editing in state \emph{2-3}. However, later on they take a more subdued approach and 
only post 
comments. The reason for this decreased interest is hard to gauge 
but identifying these users and retaining their interest could be important to sustain the community. 

\vspace{0.2in}
In summary, we identify four archetypes for researchers: steady, diverse, evolving and diffuse. We observe differences in evolution of male and female researchers within the same archetype. When we examine the diverse archetype in detail, we observe that women and men differ in how they start, how they transition and time spent in mid-career. The differences in grant income are salient across states within an archetype. We also observe differences across genders within a stage of an archetype. We also identify archetypes for StackOverflow: experts, information seekers, enthusiasts and facilitators.


\begin{table}
	\centering
	\small
	\begin{tabular}{@{}l p{15mm} p{15mm} p{15mm} p{15mm}@{}}
		\toprule
		  & Experts & Seekers & Enthusiasts & Facilitators \\
		\midrule
		State 1& \{14.4S\} \newline $\mathbf{A}$ (68\%) \newline C (19\%) &
		\{3.7S\} \newline $\mathbf{Q}$ (69\%)\newline C (19\%)&
		\{12.6S\} \newline $\mathbf{C}$ (42\%)\newline \textbf{Q} (39\%) &
		\{15.4S\} \newline $\mathbf{Q}$(50\%) \newline C (32\%) \\
		State 2 &  \{17.8S\} \newline \textbf{C} (60\%)\newline ${A}$ (25\%) &
		\{8.4S\} \newline \textbf{C} (51\%)\newline ${Q}$ (33\%) &
		\{9.5S\} \newline \textbf{A} (49\%)\newline ${C}$ (32\%) \newline ${E}$ (12\%)&
		\{26.4S\} \newline \textbf{C} (44\%) \newline ${A}$ (28\%) \newline ${E}$ (22\%)\\
		State 3 & \{9S\} \newline $\mathbf{A}$ (87\%)  &
		\{2.2S\} \newline $\mathbf{A}$ (84\%) &
		\{3.7S\} \newline $\mathbf{E}$ (82\%) &
		\{8.4S\} \newline $\mathbf{A}$ (87\%)  \\
		State 4 & \{12.3S\} \newline $\mathbf{C}$ (82\%)  &
		\{5.3S\} \newline $\mathbf{C}$ (72\%) \newline ${Q}$ (15\%) &
		\{9S\} \newline $\mathbf{C}$ (75\%) \newline $A$ (10\%)  &
		\{24.2S\} \newline $\mathbf{C}$ (68\%) \newline $\mathbf{A}$ (14\%) \newline $E$ (13\%) \\
		State 5 & \{5S\} \newline $\mathbf{E}$ (45\%) \newline {C} (28\%) \newline ${A}$ (21\%) &
		\{3.7S\} \newline $\mathbf{Q}$ (85\%) &
		\{10.5S\} \newline $\mathbf{Q}$ (40\%) \newline {C} (33\%) \newline ${E}$ (16\%) &
		\{14.3S\} \newline \textbf{E} (63\%) \newline ${C}$ (22\%) \\

		State 6 & \{11S\} \newline $\mathbf{C}$ (48\%) \newline {A} (38\%)  &
		\{7.7S\} \newline $\mathbf{C}$ (55\%) \newline \textbf{Q} (23\%) \newline ${E}$ (11\%) &
		\{5S\} \newline $\mathbf{A}$ (61\%) \newline {C} (23\%) \newline ${E}$ (11\%) &
		\{21S\} \newline \textbf{C} (57\%) \newline ${E}$ (24\%) \newline ${A}$ (13\%) \\
		\bottomrule
	\end{tabular}
	\caption{ \small \label{tab:stackexchangemean} Learned mean vector for each state for four archetypes in the Stack Overflow Dataset. We list the \emph{activities} in sorted order and annotate them with their \% contribution in the state. We list main activities ($>$ 11\%) for each state. Each state is also labeled with it's average number of sessions. The labels reflect our own interpretation of the user behavior.}
\end{table}

%% file: Discussion.tex
\subsection{Archetype variations across Gender}
\label{sec:discussion}
In this section, we analyze the evolution of research interests of male and female researchers. To this end, we manually annotate gender of all current and emeritus professors in top 50 Computer Science Universities as reported by U.S. News \& World Report\footnote{\url{bit.ly/usnews-cs}}. We consider only current and emeritus \emph{Full} Professors as they typically have 15 or more years of publication history. This results in a total of 1084 authors, 127 of whom are women.

\begin{figure*}[tbh]
	\centering
	\begin{subfigure}{\textwidth}
		\centering
		\includegraphics[width=\linewidth,height=7cm]{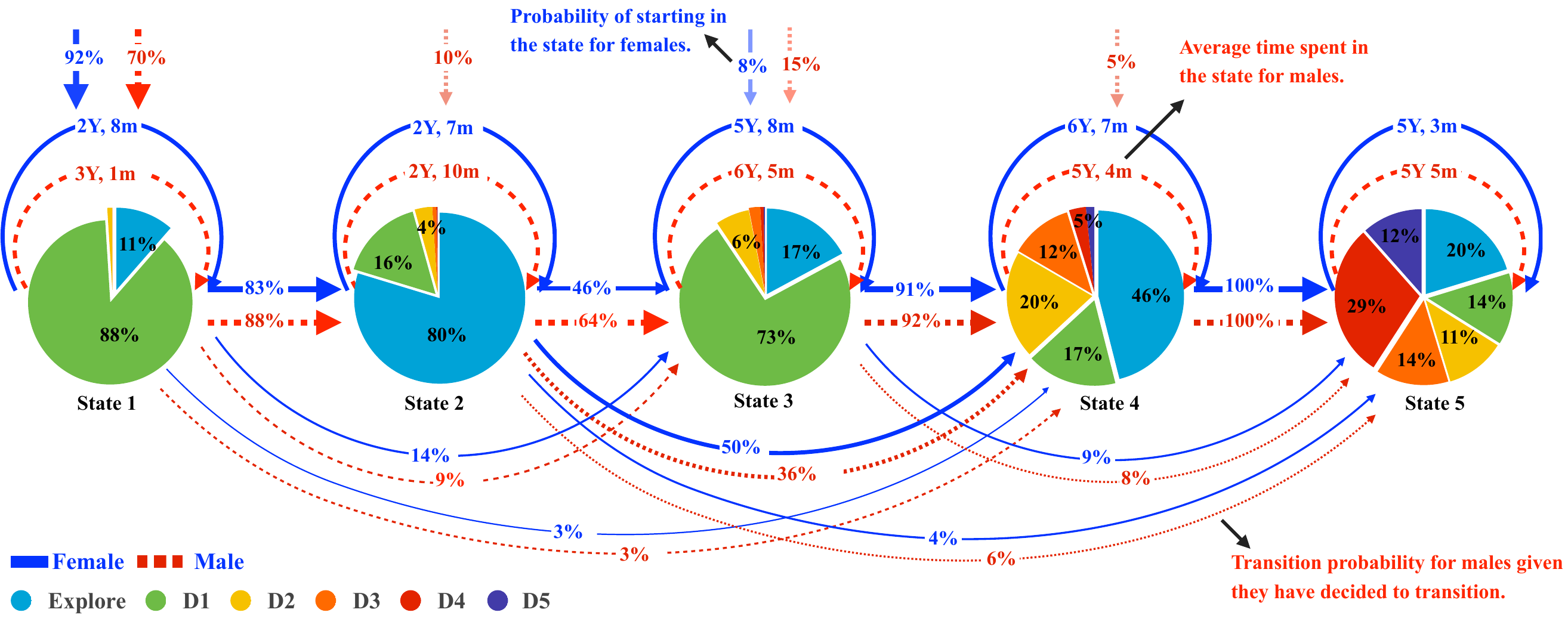}
	\end{subfigure}
	\caption{\small \label{fig:genderacadclusters} Gender wise representation of trajectory for researchers belonging to the \emph{diverse} archetype in the Academic Dataset. The transitions in blue denote transition probabilities of female professors in the archetype while those in red represents probabilities for their male counterparts. Men start their career from later evolved stages while women make long term state transitions.}
	\label{fig:academic_gender}
\end{figure*}

While researchers from both genders in the same archetype will traverse the same set of stages, they may differ in \textit{how} they transition $\tau^c$, and at \textit{which} stage they start $\pi^c$. For this analysis, we estimate separate model parameters for female $\lambda^c_f$ and male $\lambda^c_m$ researchers 
for each archetype $c$.

\begin{table}[tbh]
	\centering
	\small
	\begin{tabular}{lllll} 
		\toprule %
		Gender & Steady & Diverse & Evolving & {Diffuse} \\
		\midrule
		Male   & 2.10***  & 2.63**  & 1.15  & 1.10   \\ 
		Female & 1.80***  & 1.64**  & 1.60***  & 1.38***    \\
		\bottomrule 
	\end{tabular}
	\caption{ \small
	\label{tab:genderclusterdata} Likelihood ratio for academics across archetypes. It measures odds of a researcher being better explained by model for their gender than by model for the other gender.\\ $* = p < .05, ** = p< .01, *{*}* = p < .001$}
\end{table}

To quantify the difference between two models ($\lambda^c_f, \lambda^c_m$) for archetype $c$, we compute their \emph{likelihood ratio}. Likelihood ratio $R^c_f$ of female researchers in archetype $c$ is:

{\small 
\begin{equation}
	R^c_f = \exp \left ( \frac{1}{|N^c_f|} {\sum_{i \in {N_f}} \log \frac{ P(X_i | \lambda_f^c)}  {P(X_i | \lambda_m^c)}} \right )  
	\label{eq:confusion metric}
\end{equation}
}%

where $N^c_f$ represents all female researchers in $c$-th archetype. The equation simplifies to say that $\log R^c_f$ is the average difference between log likelihoods of a trajectory of a female researcher generated from model for female researchers with male model of the same archetype. Thus, for instance value of $R^c_f = 2$ denotes that female researchers are twice more likely to be generated by the model of their own gender than of the opposite gender. We compute a similar ratio $R^c_m$ for men

\Cref{tab:genderclusterdata} shows the likelihood ratio and $p$-value for the paired-sample t-test \citet{goulden:1949} between the likelihood values. Since most of the values are statistically significant, all researchers are better explained by the model for their gender, than by the model for the opposite gender. Male researchers are distinct for the steady and diverse archetypes, but not for the evolving and diffuse archetypes. For women, on an average, the effect is larger, with the strongest effects seen for the steady, diverse, and evolving archetypes.





For the sake of brevity, we examine gender difference in only the \emph{diverse} archetype in some detail. ~\Cref{fig:academic_gender} shows three interesting variations. First, we observe that women are much more likely to start in state 1 (92\%), with a dominant area of interest ($D_1$) than in any other state. In contrast, men start in states 1, 2, 3 and 4, with only 70\% starting in state 1. Both men and women skip stages, but women are more likely to skip a stage than men. For example, 50\% of women skip stage 3, while only 36\% men do. Longer skips of two stages are more rare, and both women and men make these long skips at the same rate. Finally, there are clear differences between mid-career men and women (states 3, 4): women spend more time \emph{exploring} mid-career (state 4) than men, and mid-career men spend more time in their starting area of interest ($D_1$, state 3) than women.

\subsection{Grant income variability across Archetypes \& Gender}
\label{sec:grant}
\begin{figure}[tbh]
	\centering
	\begin{subfigure}{0.25\textwidth}
		\centering
		\includegraphics[width=\linewidth,height=4cm]{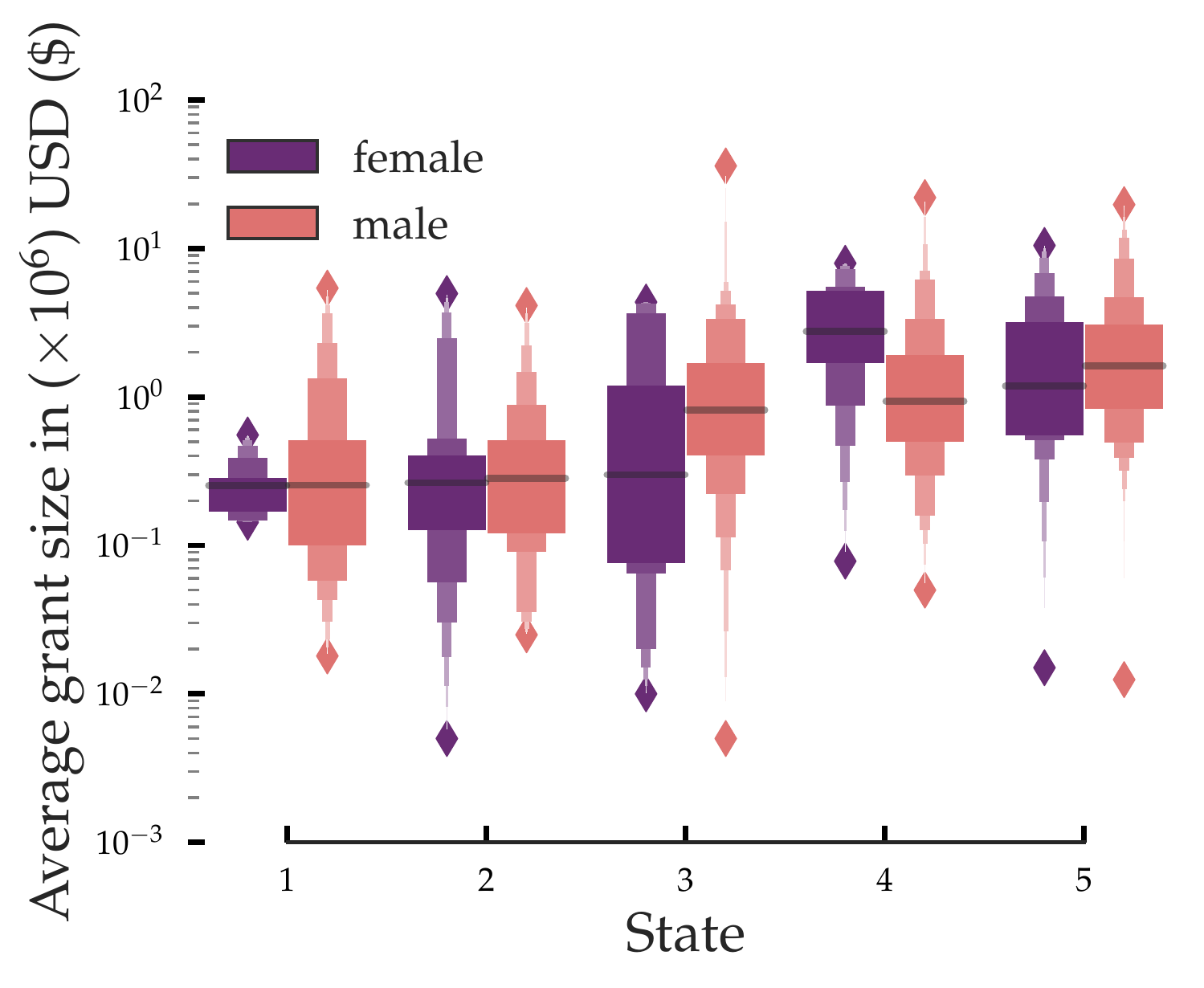}
		\caption{Steady Researchers}
	\end{subfigure}%
	\begin{subfigure}{0.25\textwidth}
		\centering
		\includegraphics[width=\linewidth,height=4cm]{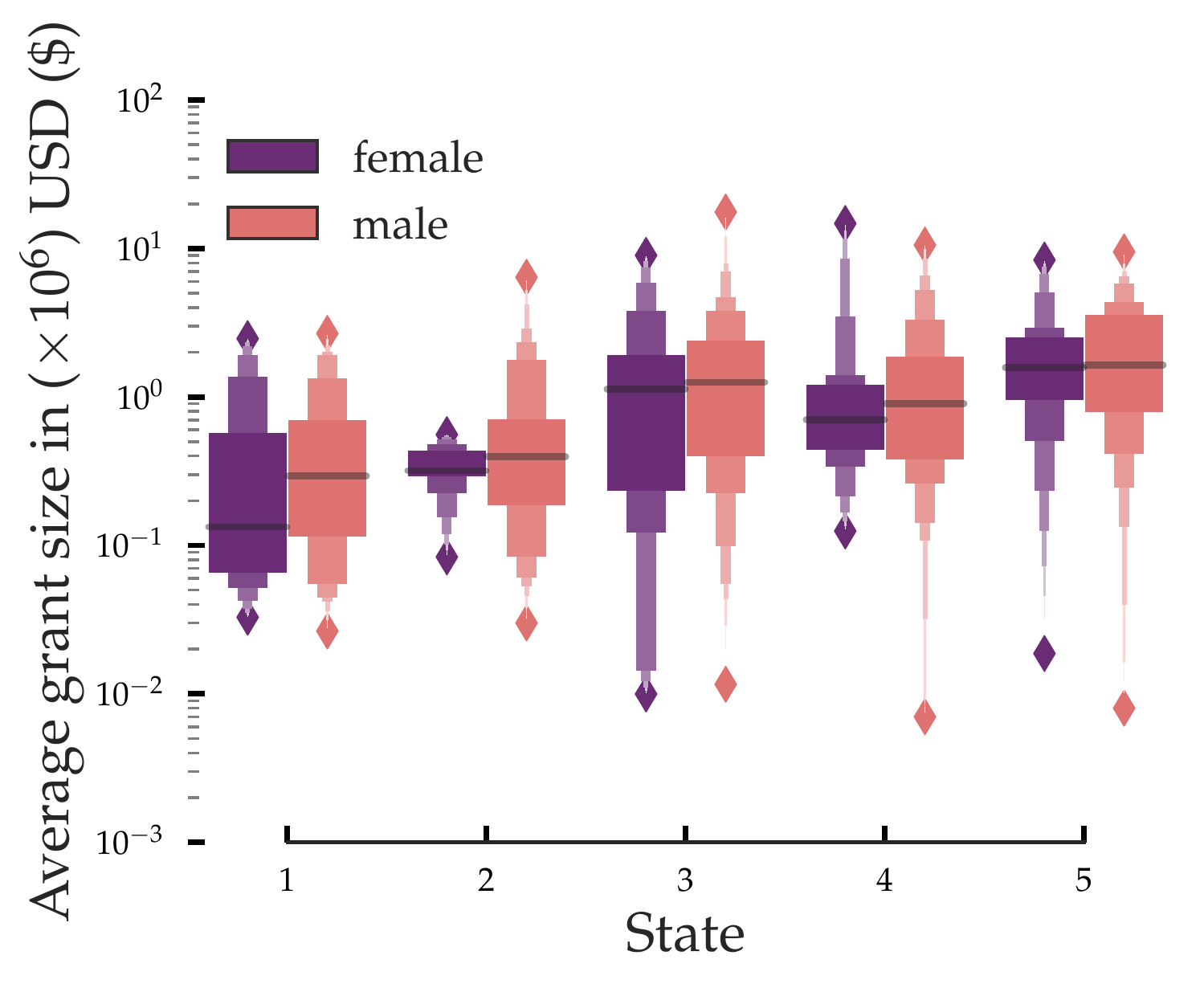}
		\caption{Diverse Researchers}
	\end{subfigure}
	\begin{subfigure}{0.25\textwidth}
		\centering
		\includegraphics[width=\linewidth,height=4cm]{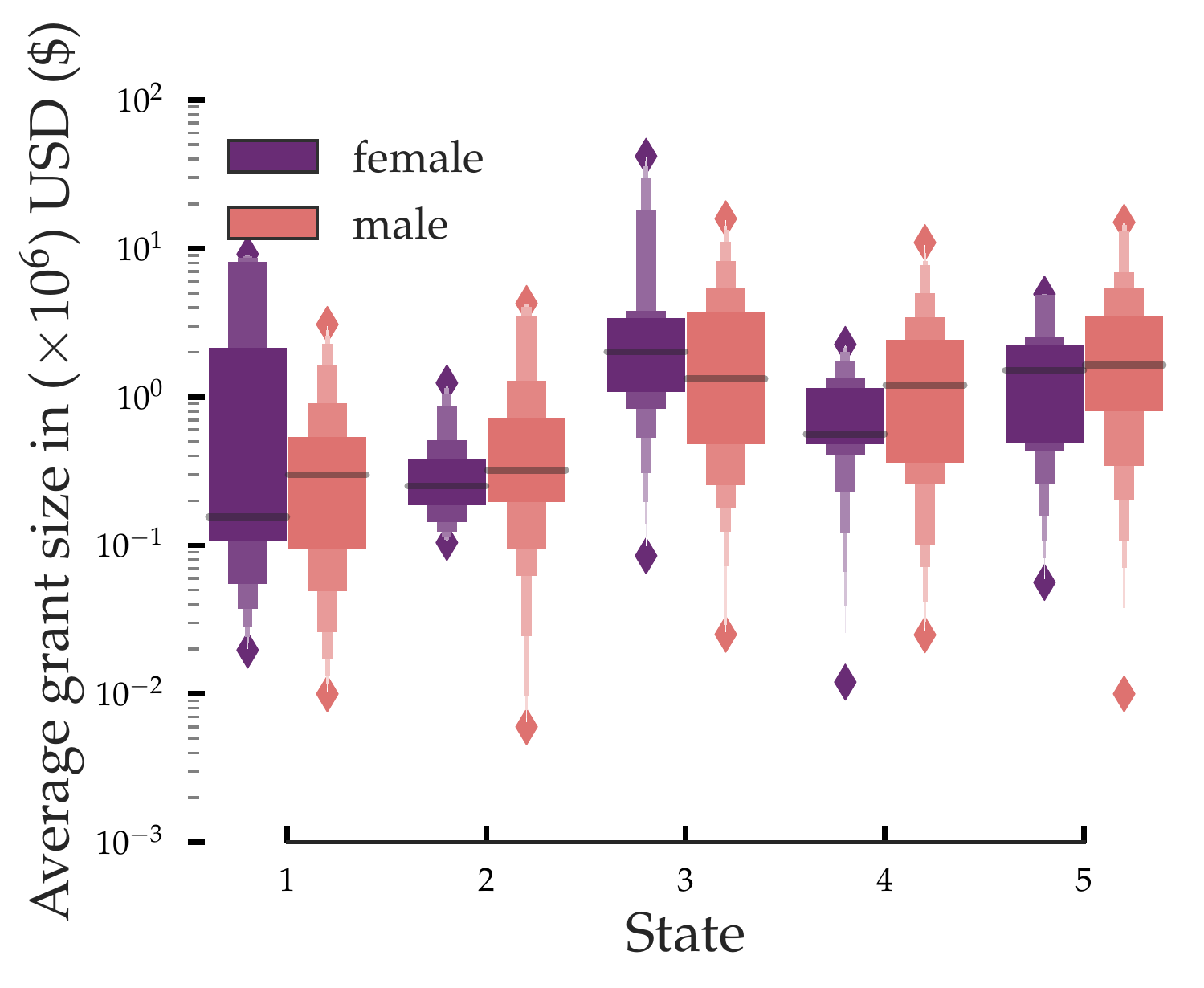}
		\caption{Evolving Researchers}
	\end{subfigure}%
	\begin{subfigure}{0.25\textwidth}
		\centering
		\includegraphics[width=\linewidth,height=4cm]{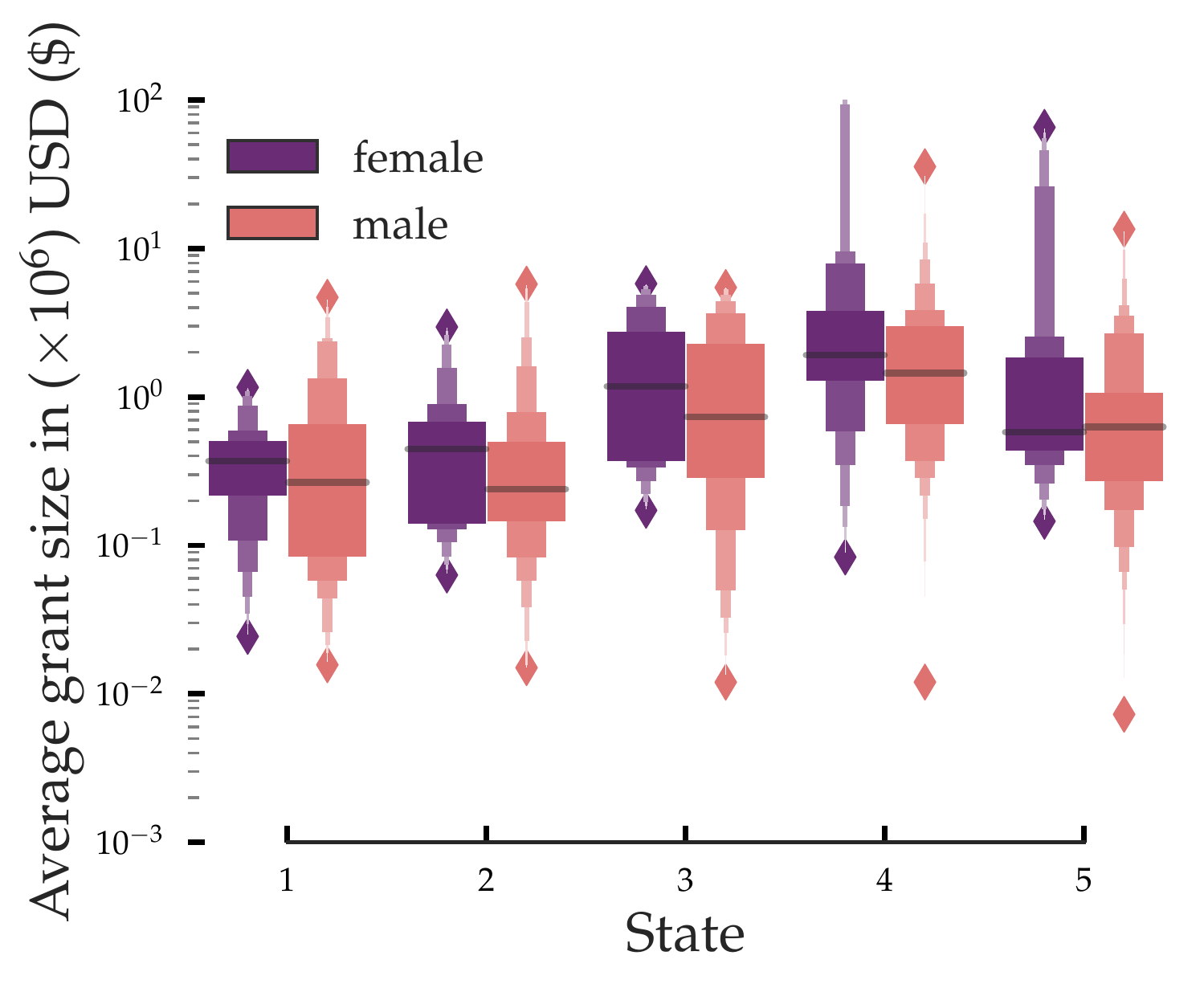}
		\caption{Diffusive Researchers}
	\end{subfigure}
	\caption{\small \label{fig:grantanalysis} Letter value plots of total grant money awarded by NSF when author is a PI in each stage. In general, Professors get more grant money as they gain experience. Regardless of archetypes, grant income in state 3 is significantly higher from state 2 (p< .01). There are also significant differences across genders within a state of an archetype. For instance, for Evolving archetype, male professors get significantly more income than female professors in state 4 (p < .01).}
\end{figure}

Now, we examine the relationship between variation in the academic trajectories and gender to research grants awarded at different stages of academic career. We extract historical information of grants from the National Science Foundation, a large federal funding agency for Science \& Engineering in the United States \footnote{\url{bit.ly/nsfgrants}}. A PI \emph{leads} the grant, while a Co-PI \emph{collaborates} in the grant. We analyze the same subset of CS professors in top-50 US universities as in~\Cref{sec:discussion}. We collect information for 1062 professors and manually disambiguate names and identify gender by cross-validating with the researcher's 
webpage. 
Then, we compute the average grant money awarded to a researcher, at each stage in their trajectory.~\Cref{fig:grantanalysis}, which shows letter-value plots of average grant size awarded as PI's, broken down by archetypes (steady, diverse, evolving or diffuse), stage within an archetype and gender, summarizes our findings.


\begin{table}[tbh]
	\centering
	\small
	\begin{tabular}{ll} 
		\toprule %
		Archetype (H-test) &  State Pair (t-test)  \\
		\midrule
		\multirow{1}{*}{Steady***}    & State 2 vs 3**    \\ 
		   			                \hline
		\multirow{2}{*}{Diverse***}   & State 2 vs 3*** \\ 
						        	& State 4 vs 5*    \\ \hline 
		\multirow{3}{*}{Evolving***}  & State 2 vs 3***   \\ 
						 & State 3 vs 4*    \\ 
						& State 4 vs 5**    \\ \hline 
		\multirow{2}{*}{Diffuse***}   & State 2 vs 3*** 	\\ 
		             & State 4 vs 5*    \\ 
		\bottomrule 
	\end{tabular}
	\caption{ \small
		Statistical significance tests for the differences in grant money across latent states within an archetype. Shown are only those tests that are statistically significant. H-test\cite{Kruskal:1952} confirms that at least one state is different from another state of the archetype; t-test\cite{Welch:1947} was then conducted between each consecutive states within the archetype to determine the differing states.  \\ $* = p < .05, ** = p< .01, *{*}* = p < .001$ }
	\label{tab:statstate}
\end{table}

\begin{table}[tbh]
	\centering
	\small
	\begin{tabular}{ll} 
		\toprule %
		Archetype & Latent State (t-test) \\
		\midrule
		\multirow{2}{*}{Steady}   & State 1* \\ 
							& State 4*  \\ \hline 
		Diverse   &  State 2*  \\ \hline 
		\multirow{2}{*}{Evolving}  & State 4** \\ 
		 					& State 5*    \\  \hline 
		Diffuse & {Not significant}  \\
		\bottomrule 
	\end{tabular}
	\caption{ \small
	\label{tab:statgender}Statistical significance tests \cite{Welch:1947} for the differences in grant money across gender in each state within an archetype. Shown are only those tests that are statistically significant. \\ $* = p < .05, ** = p< .01, *{*}* = p < .001$}
\end{table}

Additionally, we conducted Kruskal-Wallis H-test~\citet{Kruskal:1952} to establish statistical significance of differences in grant money across latent states within an archetype. This test affirms that at least one latent state is different from another latent state within an archetype. We then conducted Welch's t-test~\citet{Welch:1947} between consecutive states to find the exact pair of states which are  significantly different. We only tested with consecutive latent states as we are only interested in grant income changes as author progresses through stages. ~\Cref{tab:statstate} reports the state pairs for each archetype that are statistically different. In the rest of this section, we describe these results in detail.

Regardless of archetypes, we observe that in general authors tend to receive more grant money as they gain experience in ~\Cref{fig:grantanalysis}. On average, across archetypes and gender, PI's receive in state 5, four times the amount of grant money than state 1 ($p < .001$). Also for researchers across archetypes and across genders, we notice an uptick in grant income in state 3 from state 2 ($p < .01$ - ~\Cref{tab:statstate}). Let us qualitatively examine the \emph{steady} researchers in detail, by comparing~\Cref{fig:grantanalysis} with~\Cref{fig:acadclusters}. State 2 in~\Cref{fig:acadclusters} shows the researchers exploring different topics, whereas in state 3, they are spending a significant part of their time on their main domain $D_1$. Also notice that 36\% of the researchers never visit state 2 - 27\% skip state 2, and 9\% of the researchers start in state 3. Since state 1 typically represents the time spent by the researchers in their PhD, and with 74\% time spent in an explore stage in state 2, it is not surprising that we see limited grant income in their first two states. State 3, perhaps reflects a sustained focus on their domain $D_1$, and this pays off in terms of grant income. Similar qualitative arguments follow for the other archetypes.

The grant trajectories over states is different for each archetype ($p < .001$). Let us examine statistically different state pairs from ~\Cref{tab:statstate} in ~\Cref{fig:grantanalysis}. Steady researchers see a big uptick in their grant income in state 3 and their subsequent grant income is similar in magnitude ($p < .01$). The grant income for diverse researchers (who have more than one dominant area) increases steadily over states($p < .05$). For evolving researchers (who change their dominant area), the grant income rises (state 3, $p < .001$), falls (state 4, $p < .05$) and rises (state 5,$ p < .01$), reflecting a degree of unpredictability accompanying changing area of interest. Diffuse researchers, have a pattern similar to steady researchers except in state 5 ($p < .05$), when the income dips, perhaps due to spending time in too many areas.

To determine differences in grant income across gender, we conducted t-test \citet{Welch:1947} for each state within an archetype. ~\Cref{tab:statgender} reports significantly different states within each archetype. We again examine these statistically different states from ~\Cref{tab:statgender} in ~\Cref{fig:grantanalysis}. Evolving women receive significantly \emph{lower} income than evolving men when they \emph{switch to new areas} in state 4 and 5 ($p < .05$). On the other hand, in our dataset, steady women receive significantly \emph{higher} grant income than steady men when they \emph{switch areas} in state 4 (p $<$ .05). In general, we observe that men show greater grant income variability than do women. The variability is statistically significant ($p < .05$) during early career in state 1 and 2 for Steady and Diverse researchers respectively. We do not observe significant differences in grant income of male and female Diffusive researchers.

%% file: limitations.tex

\section{Limitations}
\label{sec:Limitations}

Our proposed model identified insightful archetypes and its variability with gender and grant income of professors. However, it is important to understand certain caveats to the reported findings.
First, in terms of the data, the discovered archetypes for academics are for the top researchers in their field (we pick \emph{influential} researchers in each of the 35 research subdomains). Thus, our archetypes do not reflect all computer scientists engaged in research. Also our grant analysis was focused on professors from only top-50 Computer Science schools. In our current study, we collected grant history from data publicly available by NSF. The funding analysis can be extended by collecting data from other possible funding sources like National Institute of Health (NIH), gifts and professor's salary. Hence, we believe that our study is a first step in understanding differences in research conducting behavior of academics and its effect on their income.

Second, as with all inductive models, our qualitative results depend on chosen model. Recently, Deep Neural Networks, especially Recurrent Neural Networks have been proposed to model time series data. There has also been great interest in building interpretable models \citep{Ribeiro:2016, hima:2016}. However, still most of these models remain as a black box and do not provide meaningful results. 

Third, in our current version of the model, we do not consider the effect of collaborations, or the role of conferences where researchers publish, and where they may pick up on normative behavior (e.g. areas in which to work) on the discovered archetype.In future work, we plan to understand the role of community interaction on archetypes and address these limitations. Another interesting research direction is to explore correlation of change in research behavior with career transitions and author's citation count.


%% file: Conclusion.tex
\section{Conclusion}
\label{sec:conclusion}
In this paper, we aimed to discover archetypical behavioral patterns for individuals in large social networks. The observation that despite near limitless variation in behavior, the change in behavior exhibits regularities, motivated our research. We introduced a novel Gaussian Hidden Markov Model Cluster (G-HMM) to identify archetypes and evolutionary patterns within each archetype. We chose to work with G-HMM's since they allow for : near limitless variation; constraints how individuals can evolve; different evolutionary rates and are parsimonious.

We identified four archetypes for computer scientists: steady, diverse, evolving and diffuse and showed examples of computer scientists from different sub-fields that share the same archetype. We analyzed full professors from the top 50 CS departments to understand gender differences within archetypes.
Women and men differ within an archetype (e.g. diverse) in where they start, rate of transition and research interests during mid-career. We further analyzed grant income of these professors to understand the effect of gender and archetype on income. The differences in income are salient across states within an archetype. There also exist significant differences across genders within a state of an archetype. For StackOverflow, discovered archetypes could be labeled as: \emph{Experts}, \emph{Seekers}, \emph{Enthusiasts} and \emph{Facilitators}. We showed strong quantitative results with competing baselines for future activity prediction and perplexity.